\documentclass[12pt]{article}
\usepackage{epsfig} 
\def\beq{\begin{equation}}
\def\eeq{\end{equation}}

\def\al{\alpha}
\def\bt{\beta}

\def\ga{\gamma}
\def\de{\delta}
\def\De{\Delta}

\def\ka{\kappa}

\def\te{\theta}

\def\lam{\lambda}

\def\om{\omega}
\def\ep{\epsilon}

\def\sq{\sqrt}

\def\l{\left (}
\def\r{\right )}
\def\lq{\left [}
\def\rq{\right ]}  
\def\fr{\frac}
\def\la{\label}
\def\hs{\hspace}
\def\vs{\vspace}

\def\ran{\rangle}
\def\lan{\langle}
\def\ov{\overline}
\def\tl{\tilde}
\def\tm{\times}

\begin{document}

\begin{flushright}
BA-04-01\\
HD-THEP-04-04\\
January 27, 2004 \\
\end{flushright}

\begin{center}
{\Large\bf    

Suppressed $\te_{13}$ In A Democratic Approach}
\end{center}

\vspace{0.5cm}
\begin{center}
{\large 
{}~Qaisar Shafi$^{a}$\footnote{E-mail address: 
shafi@bxclu.bartol.udel.edu}~~and~ 
{}~Zurab Tavartkiladze$^{b}$\footnote{E-mail address: 
Z.Tavartkiladze@ThPhys.Uni-Heidelberg.DE} 
}
\vspace{0.5cm}

$^a${\em Bartol Research Institute, University of Delaware,
Newark, DE 19716, USA \\

$^b$ Institute for Theoretical Physics, Heidelberg University,
Philosophenweg 16, \\
D-69120 Heidelberg, Germany}


\end{center}
\vspace{0.6cm}

\begin{abstract}

Within a democratic approach based on discrete symmetries, we show how the
charged lepton mass hierarchies and bilarge neutrino mixings can be
realized. For the third mixing angle we find
$\te_{13}\sim \sq{\fr{\De m_{\rm sol}^2}{\De m_{\rm
atm}^2}}\sim 0.1-0.2$.

\end{abstract}


\vs{0.5cm}

\section{Introduction}

Recent data has provided increasing evidence for atmospheric \cite{atm}
and solar 
\cite{sol}, \cite{KamL} neutrino oscillations, suggesting neutrino
masses such that $\De m_{\rm atm}^2\simeq 2\cdot 10^{-3}~{\rm eV}^2$,
$\De m_{\rm sol}^2\simeq 7\cdot 10^{-5}~{\rm eV}^2$, and
bilarge neutrino mixings:
$\sin^2 2\te_{\mu \tau }\simeq 1$, 
$\sin^2 2\te_{e \mu ,\tau }\approx 0.84$. 
It is certainly challenging to provide a self consistent theoretical
explanation of  these mixing angles and
mass hierarchies (including the charged fermions). 
The introduction of flavor symmetries which distinguish
the different generations seems to be one reasonable approach.
An alternative idea is the democratic approach, according to which all
entries in the appropriate Yukawa matrices are of the same order of
magnitude.
Applying this type of construction to the
neutrino sector \cite{demo1}-\cite{demo2}, one can naturally get large
angles in the lepton mixing matrix. However, with neutrino democracy
one also expects the third mixing angle  $\te_{13}$ to be large. On
the other hand, CHOOZ experiment \cite{chooz} provides the upper 
bound $\te_{13}\stackrel{<}{_\sim }0.2$,
and future long-baseline experiments could access $\te_{13}$ even down
to $\sim 10^{-2}$ \cite{te13}, which should put 
severe constraints on  model building.

It is clearly desirable to have an (elegant) explanation of 
how $\te_{13}$ is small  within neutrino democracy.
The origin of the ratio 
$\fr{\De m_{\rm sol}^2}{\De m_{\rm atm}^2}=0.017-0.053$ and the
hierarchies between charged lepton masses also should be simultaneously
understood in this approach.

The aim of this paper is to gain an understanding of the issues
mentioned above within the MSSM framework.
We present a democratic scenario based on the 
permutation symmetry $S_2^l\tm S_3^{e^c}$ acting in the lepton
sector. Starting with the charged lepton sector, we show how this
symmetry, especially its breaking, can be exploited to yield an
explanation of the small ratios
$\fr{m_{\mu }}{m_{\tau }}$, $\fr{m_e}{m_{\tau }}$.
Next we extend our studies to the neutrino sector and demonstrate how
the large atmospheric and solar neutrino oscillations are realized. 
Finally, we turn to $\te_{13}$ and show that the model predicts a
suppressed (but not too small) value of\footnote{A similar relation 
has previously been discussed in \cite{akhm}. We thank E. Akhmedov for
bringing it to our attention. Employing discrete symmetries for realizing 
bilarge mixings and a small $\te_{13}$ has been advocated by several
authors. See, for instance, \cite{babu}.} 
$\te_{13}\sim \sq{\fr{\De m_{\rm sol}^2}{\De m_{\rm
atm}^2}}=0.13-0.23$. These values satisfy the CHOOZ experimental bound and
hold promise that the model will be tested in the near future \cite{te13}.

\section{Charged Lepton Sector}

We start our discussion with the charged lepton sector and the
symmetries which will play a crucial role in generating natural mass
hierarchies 
and desirable lepton mixing angles. In the left-handed lepton sector, we 
introduce the symmetry $S_2^l$ which exchanges $l_1$ and $l_3$ states
and leaves $l_2$ intact:
\beq
S^l_2~:~~~l_1\to l_3~,~~~l_3\to l_1~,~~~l_2\to l_2~.
\la{Sl2}
\eeq
This symmetry it turns out guarantees the smallness of  $\te_{13}$.
Indeed, in
the limit when the solar neutrino mass scale is neglected, 
$\te_{13}$ turns out to be zero.

{}For the right handed $e^c$ states we introduce the permutation 
symmetry $S_3^{e^c}$ which acts as follows:
\beq
S_3^{e^c}~:~~~~e_1^c\to e_2^c~,~~~~
e_2^c\to e_3^c~,~~~~e_3^c\to e_1^c~.
\la{Se3}
\eeq
It turns out that in the limit of unbroken $S_2^l\tm S_3^{e^c}$,
only a single charged lepton (tau) acquires mass. The electron 
and muon acquire masses from $S_2^l\tm S_3^{e^c}$ breaking. 
Therefore, the hierarchies $m_e/m_{\tau }$, $m_{\mu }/m_{\tau }$
can be controlled by the symmetry breaking pattern.

To break $S_2^l\tm S_3^{e^c}$ we introduce SM singlet superfields $X^l$ and 
${\bf X}^{e^c}=(X_1^{e^c},~X_2^{e^c},~X_3^{e^c})$
which under $S_2^l$ and $S_3^{e^c}$ have the following transformation 
properties 
\beq
S^l_2~:~~~X^l\to -X^l~,~~~~
\la{XN}
\eeq
\beq
S_3^{e^c}~:~~~~X^{e^c}_1\to X^{e^c}_2~,~~~~
X^{e^c}_2\to X^{e^c}_3~,~~~~
X^{e^c}_3\to X^{e^c}_1~.
\la{SX3}
\eeq
We now write down the Yukawa interactions which are responsible for
charged lepton masses and respect $S_2^l\tm S_3^{e^c}$
symmetry. Those couplings that do not involve the fields
$X^l$, $X_{1,2,3}^{e^c}$ and give
dominant contribution to the charged lepton mass matrix are
\begin{equation}
\begin{array}{ccc}
 & {\begin{array}{ccc}
\hs{-0.1cm} l_1&\hs{0.22cm} l_2&\hs{0.2cm} l_3\hs{0.8cm}
\end{array}}\\ \vspace{2mm}

\begin{array}{c}
 e^c_1\\ e^c_2 \\ e^c_3  

\end{array}\!\!\!\!\!\! &{\left(\begin{array}{ccc}
1~ &~ \rho ~&~1
\\
1~ &~ \rho ~&~1 
\\
1~ &~ \rho ~&~1 
\end{array}\!\right )}\lam_Eh_d~,
\end{array}  
\label{leading}
\end{equation} 
where $h_d$ is the down type Higgs doublet superfield and
$\rho $, $\lam_E$ are dimensionless couplings. Due to democracy
$\rho $ is of order  unity, while $\lam_E$ determines the value of the
MSSM parameter $\tan \bt $. 
The couplings involving $X_{1,2,3}^{e^c}$ and $X^l$  are
respectively
$$
\fr{\lam_Eh_d}{M}\lq X^{e^c}_1(p e^c_1+q e^c_2+r e^c_3)+
X^{e^c}_2(r e^c_1+p e^c_2+q e^c_3)+
X^{e^c}_3(q e^c_1+r e^c_2+p e^c_3)\rq (l_1+l_3)+
$$
\beq
\fr{\lam_Eh_d}{M}\lq X^{e^c}_1(\rho_1 e^c_1+\rho_2 e^c_2+\rho_3 e^c_3)+
X^{e^c}_2(\rho_3 e^c_1+\rho_1 e^c_2+\rho_2 e^c_3)+
X^{e^c}_3(\rho_2 e^c_1+\rho_3 e^c_2+\rho_1 e^c_3)\rq l_2~,
\la{subleadR}
\eeq
\begin{equation}
\begin{array}{ccc}
 & {\begin{array}{ccc}
\hs{-0cm} l_1&\hs{0.4cm} l_2&\hs{0.5cm} l_3\hs{1.5cm}
\end{array}}\\ \vspace{2mm}

\begin{array}{c}
 e^c_1\\ e^c_2 \\ e^c_3  

\end{array}\!\!\!\!\!\! &{\left(\begin{array}{ccc}
1~ &~ 0 ~&~-1
\\
1~ &~ 0 ~&~ -1
\\
1~ &~ 0 ~&~-1
\end{array}\!\right )}\fr{X^l}{M}\lam_Eh_d~,
\end{array}  
\label{subleadL}
\end{equation} 
where $M$ is some cut off scale and
$p, q, r, \rho_i$
are dimensionless couplings of order unity. 

Next, we
assume the following VEVs for the scalar components of the singlet
superfields:
\beq
\fr{\lan X^l\ran }{M}\equiv \ep_L\ll 1~,
\la{VEVXN}
\eeq
\beq
\fr{\lan X^{e^c}_1\ran }{M}\equiv \ep_R\ll 1~,~~~~
\lan X^{e^c}_2\ran =\lan X^{e^c}_3\ran =0~.
\la{Xe3VEV}
\eeq
According to the breaking pattern
(\ref{VEVXN}), (\ref{Xe3VEV}), the couplings in which participate 
$X_{2,3}^{e^c}$, 
are not relevant for the mass matrix. The charged 
lepton mass matrix can be written as
\begin{equation} 
\begin{array}{ccc}
 & {\begin{array}{ccc}
\hs{-1.5cm} e_1&\hs{1.7cm} e_2~~~&\hs{1.5cm} e_3
\end{array}}\\ \vspace{2mm}
m_E=
\begin{array}{c}
 e^c_1\\ e^c_2 \\ e^c_3  

\end{array}\!\!\!\!\!\! &{\left(\begin{array}{ccc}
1+p\ep_R+\ep_L~ & \rho +\rho_1\ep_R~ &1+p \ep_R-\ep_L
\\
1+q\ep_R+\ep_L~ & \rho +\rho_2\ep_R~ & 1+q\ep_R-\ep_L
\\
1+r\ep_R+\ep_L~ & \rho +\rho_3\ep_R~ & 1+r\ep_R-\ep_L
\end{array}\!\right )}\fr{\ov{m}}{\sq{3(2+\left |\rho \right |^2)}}~,
\end{array}  
\label{mE} 
\end{equation} 
where $\ov{m}=\lam_E\sq{3(2+\left |\rho \right |^2)}\lan h_d^{(0)}\ran $ 
and a suitable normalization has been 
chosen. For analysis, it is convenient to write (\ref{mE}) as
\beq
m_E=\fr{\ov{m}}{\sq{3(2+\left |\rho \right |^2)}}
\l Y_0+\ep_RY_R+\ep_LY_L \r~,
\la{mEexp}
\eeq
where
\begin{equation}
\begin{array}{ccc}
 & {\begin{array}{ccc}
~ & &\,\,~~~
\end{array}}\\ \vspace{2mm}
\begin{array}{c}
 \\  \\ 
\end{array}\!\!\!\!\! &Y_0={\left(\begin{array}{ccc}
1 ~&\rho ~&1
\\
1 ~&\rho ~&1
\\
1 ~&\rho ~&1
\end{array}\right)~, }~
\end{array}  
\begin{array}{ccc}
 & {\begin{array}{ccc}
~ & &\,\,~~~
\end{array}}\\ \vspace{2mm}
\begin{array}{c}
 \\  \\ 

\end{array}\!\!\!\!\! &Y_R={\left(\begin{array}{ccc}
p ~& \rho_1 ~& p
\\
q ~& \rho_2 ~& q
\\
r ~& \rho_3 ~& r
\end{array}\right) }~,~~
\end{array}  
\hs{-1cm}
\begin{array}{ccc}
& {\begin{array}{ccc}
 &  & 
\end{array}}\\ \vspace{2mm}
\begin{array}{c}
  \\  \\
\end{array} &Y_L={\left(\begin{array}{ccc}
1~ & 0~ & -1
\\
1~ & 0~ & -1
\\
1~ & 0~ & -1
\end{array}\right)~.
}
\end{array}
\label{Y0RL}
\end{equation}
The term proportional to $Y_0$ in (\ref{mEexp}) provides the leading
contribution in $m_E$ and is 
responsible for the tau mass $m_{\tau }$. Assuming that 
$\ep_L, \ep_R\ll 1$, for the three eigenvalues of $m_E$ matrix, we find
\beq
m_{\tau }\simeq \ov{m}~,~~~m_{\mu }\sim \ov{m}\ep_R~,~~~
m_e\sim \ov{m}\ep_R\ep_L~. 
\la{chmass}
\eeq
Thus, $\ep_R$ and $\ep_L$ are responsible for the muon and electron 
masses\footnote{In (\ref{chmass}) some coefficients of 
order unity appear in the expressions for $m_{\mu }$ and $m_e$
which will be ignored.}
such that
\beq
\ep_R\sim \fr{m_{\mu }}{m_{\tau }}\simeq 0.06~,~~~~
\ep_L\sim \fr{m_e}{m_{\tau }\ep_R}\simeq 4.7\cdot 10^{-3}~.
\la{epRL}
\eeq

The unitary matrix $U_e$ which rotates the 
left handed charged lepton states upon diagonalization of
$m_E$, can be found by diagonalizing the matrix $m_E^{\dagger }m_E$:
\beq
(m_E^{diag})^2=U_e^{\dagger }m_E^{\dagger }m_EU_e~.
\la{diagmE}
\eeq
It is easy to see that $U_e$ is mainly determined by $Y_0$, the leading 
term in (\ref{mEexp}): 
$$
\begin{array}{ccc}
 & {\begin{array}{ccc}
~ & &\,\,~~~
\end{array}}\\ \vspace{2mm}
\begin{array}{c}
 \\  \\ 
\end{array}\!\!\!\!\! &U_e=P_{\rho }{\left(\begin{array}{ccc}
\fr{1}{\sq{2}}~& -\fr{|\rho |}{\sq{4+2|\rho |^2}}~ & 
\fr{1}{\sq{2+|\rho |^2}}
\\
0~ & \sq{\fr{2}{2+|\rho |^2}}~ & \fr{|\rho |}{\sq{2+|\rho |^2}}
\\
-\fr{1}{\sq{2}}~ & -\fr{|\rho |}{\sq{4+2|\rho |^2}}~ &
\fr{1}{\sq{2+|\rho |^2}}
\end{array}\right)+{\cal O}(\ep_R)+
{\cal O}(\ep_L)}~,
\end{array}  
\hs{-1cm}
$$
\beq
{\rm with}~~~~~~~P_{\rho }=
{\rm Diag}\l 1,~{\rm exp}[-{\rm i}\cdot {\rm Arg}(\rho )],~1 \r ~.
\label{Ue}
\eeq
The corrections of order $\sim \ep_R, \ep_L$ are
subleading and will not be further considered.
Anticipating, the (1,1) and (3,1) entries of the leading part of the neutral
lepton mixing matrix $U_{\nu }$ [see eq. (\ref{Unu0})] will have the same
values as corresponding entries of $U_e$. Because of this, there will occur
cancellations which eventually lead to the suppressed $\te_{13}$. This
effect is due
to the $S_2^l$ symmetry\footnote{The same symmetry plays a crucial role
if non canonical kinetic terms are included in the Lagrangian. 
We have checked
that all results are robust even in the presence of such
terms.}.


To summarize, with the help of $S_2^l\tm S_3^{e^c}$ symmetry,
we have gained an understanding of the hierarchies between the charged
lepton masses. 
A non trivial mixing matrix $U_e$ (\ref{Ue}) is also generated and
will contribute to the physical lepton mixing matrix.

\section{Neutrino Sector}

We introduce a single right handed neutrino $N$ which provides the
dominant contribution to the neutrino mass matrix. 
The couplings
\beq
\l \lam_Nl_1+\lam_N'l_2+\lam_Nl_3\r Nh_u-\fr{1}{2}M_NN^2~,
\la{NDirMaj}
\eeq
are invariant under $S_2^l$, and after integrating out the $N$ state,
induce an effective dimension five operator
$\fr{1}{2}lm_{\nu }^{(0)}l$, with
\begin{equation}
\begin{array}{ccc}
 & {\begin{array}{ccc}
~ & &\,\,~~~
\end{array}}\\ \vspace{2mm}
\begin{array}{c}
 \\  \\ 
\end{array}\!\!\!\!\! &m_{\nu }^{(0)}= {\left(\begin{array}{ccc}
1& t & 1
\\
t & t^2 & t
\\
1 & t & 1
\end{array}\right)\fr{\lam_N^2(h_u^0)^2}{M_N} }~,
\end{array}  
\hs{-1cm}
\label{mnu0}
\end{equation}     
where $t=\fr{\lam_N'}{\lam_N}$. The Yukawa couplings $\lam_N$, $\lam_N'$
are expected to be of the same order of magnitude and therefore $t$ is
of order unity.

Further, we include non renormalizable operators which provide subleading 
contribution to the neutrino mass matrix and, as we will see, are 
responsible for solar neutrino oscillations. 
The use of non-renormalizable operators for generating neutrino masses
and mixings has been discussed in the past (for instance, see \cite{laz}).
The couplings respecting $S_2^l$ symmetry are
\begin{equation}
\begin{array}{ccc}
 & {\begin{array}{ccc}
\hs{-1cm}l_1 &l_2  &l_3\,\,~~~
\end{array}}\\ \vspace{2mm}
\begin{array}{c}
l_1 \\l_2  \\l_3 

\end{array}\!\!\!\!\! &{\left(\hs{-0.2cm}\begin{array}{ccc}
~1&~\al &~\ga 
\\
~\al & ~\bt &~ \al
\\
~\ga &~ \al &~1
\end{array}\hs{-0.15cm}\right)\fr{h_u^2}{2M_1} }~,~~~~~~
\end{array}  
\hs{0.7cm}
\begin{array}{cc}
& {\begin{array}{ccc}
\hs{-1.2cm}l_1~~~ &~l_2~  &~~~l_3
\end{array}}\\ \vspace{2mm}
\begin{array}{c}
 l_1 \\l_2\\ l_3  

\end{array} &{\hs{-0.4cm}\left(\begin{array}{ccc}
1&~~~\ov{\al }&~~~~0
\\
\ov{\al }&~~~0&~-\ov{\al }
\\
0&~-\ov{\al }&~-1
\end{array}\hs{-0.2cm}\right)\fr{h_u^2X^l}{2M_2^2}~,
}
\end{array}
\label{nonren}
\end{equation}     
where $M_1$, $M_2$ are some cut off scales  related to
lepton number violation\footnote{These operators may emerge
from the decoupling of heavy states with masses $M_{1,2}$.}. 
We assume that 
\beq
M_1\sim \fr{M_2^2}{\lan X^l\ran }\sim \fr{M_N}{\ep }~,~~~~\ep \sim 0.1~.
\la{scales}
\eeq 
With this, the neutrino masses will have the hierarchical structure
$m_3>m_1, m_2$ and the value of $\ep $ in (\ref{scales}), as we will see,
is dictated from the observed value of 
$\sq{\fr{\De m_{\rm sol}^2}{\De m_{\rm atm}^2}}$. Since the dominant part
(\ref{mnu0}) of $m_{\nu }$ is responsible for 
$\De m_{\rm atm}^2\simeq 2,5\cdot 10^{-3}~{\rm eV}^2$,  for
$\lam_N \sim t\sim 1$ we estimate
$M_N/(\sin^2\bt )\simeq  1.8\cdot 10^{15}$~GeV. With all this
and $\lan X^l\ran =M\ep_L\sim M_{\rm Pl}\ep_L$, taking into
account (\ref{scales}), one has 
$M_1/(\sin^2\bt )\sim M_2/(\sin \bt )\sim 10^{16}$~GeV.

{}From (\ref{mnu0}), (\ref{nonren}), the mass matrix for the 
light neutrinos can be written as
\begin{equation}
\begin{array}{ccc}
 & {\begin{array}{ccc}
\hs{-1cm} &  &\,\,~~~
\end{array}}\\ \vspace{2mm}
\!\!\!\!\! &{m_{\nu }=\left(\hs{-0.2cm}\begin{array}{ccc}
~1&~t &~1 
\\
~t & ~t^2 &~ t
\\
~1 &~ t &~1
\end{array}\hs{-0.15cm}\right)\fr{m_0}{2+|t|^2}~+ }
\end{array}  
\hs{-0.5cm}
\begin{array}{cc}
& {\begin{array}{ccc}
\hs{-1.2cm}~ &~  &~
\end{array}}\\ \vspace{2mm}
\begin{array}{c}
  \\ \\ 

\end{array} &{\hs{-0.4cm}\left(\begin{array}{ccc}
1&~~~\al &~~~~\ga
\\
\al &~~~\bt &~\al 
\\
\ga &~\al &~1 
\end{array}\hs{-0.2cm}\right)\fr{m_0\ep }{2+|t|^2} + ~
}
\end{array}
\hs{-0.5cm}
\begin{array}{cc}
& {\begin{array}{ccc}
\hs{-1.2cm}~ &~  &~
\end{array}}\\ \vspace{2mm}
\begin{array}{c}
  \\ \\ 

\end{array} &{\hs{-0.4cm}\left(\begin{array}{ccc}
1&~~~\ov{\al }&~~~~0
\\
\ov{\al }&~~~0&~-\ov{\al }
\\
0 &~-\ov{\al }&~ -1 
\end{array}\hs{-0.2cm}\right)\fr{m_0\de \ep}{2+|t|^2}  ~,
}
\end{array}
\label{totenu0}
\end{equation}     
where
\beq
m_0=\fr{\lam_N^2(h_u^{(0)})^2}{M_N}(2+|t|^2)~,~~~
\ep =\fr{M_N}{\lam_N^2M_1}~,~~~
\de =\fr{M_1\lan X^l\ran }{M_2^2}~.
\la{defs}
\eeq
Making the shift $t\to t-\al \ep $ and then rescaling
$\ep \to (2+|t|^2)\ep $, the neutrino mass matrix reads
$$
m_{\nu }=m_{\nu }^{(0)}+m_{\nu }^{(1)}~,
$$
\begin{equation}
\begin{array}{ccc}
 & {\begin{array}{ccc}
\hs{-1cm} &  &\,\,~~~
\end{array}}\\ \vspace{2mm}
\!\!\!\!\! &{\rm with}~~~~~ {m_{\nu }^{(0)}
\simeq \left(\hs{-0.2cm}\begin{array}{ccc}
~1&~t &~1 
\\
~t & ~t^2 &~ t
\\
~1 &~ t &~1
\end{array}\hs{-0.15cm}\right)\fr{m_0}{2+|t|^2}~,~~~~~ }
\end{array}  
\hs{-0.5cm}
\begin{array}{cc}
& {\begin{array}{ccc}
\hs{-1.2cm}~ &~  &~
\end{array}}\\ \vspace{2mm}
\begin{array}{c}
  \\ \\ 

\end{array} &{m_{\nu }^{(1)}=\left(\begin{array}{ccc}
1+\de &~~~\tl{\al }&~~~~\ga
\\
\tl{\al }&~~~\tl{\bt }&~-\tl{\al }
\\
\ga &~-\tl{\al }&~1-\de 
\end{array}\hs{-0.2cm}\right)m_0\ep  ~,
}
\end{array}
\label{totenu}
\end{equation}     
where
$\tl{\al }=\ov{\al }\de $~, $\tl{\bt }=\bt -2t\al +\al^2\ep $~.

The leading part 
$m_{\nu }^{(0)}$ is diagonalized by the transformation
\beq
(U_{\nu }^{(0)})^Tm_{\nu }^{(0)}U_{\nu }^{(0)}
={\rm Diag}\l 0,~0,~m_0\r ~,
\la{mnu0diag}
\eeq
where
$$
\begin{array}{ccc}
 & {\begin{array}{ccc}
 & & 
\end{array}}\\ \vspace{2mm}
U_{\nu }^{(0)}=
\begin{array}{c}
\end{array}\!\!\!\!\!\! &P_t{\left(\begin{array}{ccc}
\hs{-0.1cm}\fr{1}{\sq{2}}~,&
-\fr{|t|}{\sq{4+2|t|^2}} ~, &\fr{1}{\sq{2+|t|^2}}
\\
\hs{-0.2cm}0~, &
\hs{-0.1cm}\sq{\fr{2}{2+|t|^2}}~,& \fr{|t|}{\sq{2+|t|^2}}
\\
\hs{-0.1cm}-\fr{1}{\sq{2}}~,&-\fr{|t|}{\sq{4+2|t|^2}} ~,&
\fr{1}{\sq{2+|t|^2}}
\end{array} \right)\! }~,
\end{array}  
$$
\beq
{\rm with}~~~~~~~P_t=
{\rm Diag}\l 1,~{\rm exp}[-{\rm i}\cdot {\rm Arg}(t)],~1 \r ~.
\label{Unu0}
\end{equation} 
Note that only with this transformation, the lepton mixing 
matrix $V^l=U_e^{\dagger }U_{\nu }^{(0)}$ would yield 
$\te_{12}=\te_{13}=0$. However, the subleading term $m_{\nu }^{(1)}$
in (\ref{totenu}) can provide naturally large 1-2 mixing, with
$\te_{13}$ still remaining small.
To see this, we will perform the transformation 
$(U_{\nu }^{(0)})^Tm_{\nu }U_{\nu }^{(0)}$, under which 
with redefinitions 
$\tl{\al }\to {\rm exp}[{\rm i}\cdot {\rm Arg}(t)]\tl{\al }$,
$\tl{\bt }\to {\rm exp}[{\rm i}\cdot 2{\rm Arg}(t)]\tl{\bt }$
the neutrino mass matrix (\ref{totenu}) becomes 

\begin{equation}
\begin{array}{ccc}
 & {\begin{array}{ccc}
\hs{-1cm} &  &\,\,~~~
\end{array}}\\ \vspace{2mm}
\!\!\!\!\! &{m_{\nu }'=\left(\hs{-0.2cm}\begin{array}{ccc}
~0&~0 &~0 
\\
~0 & ~0 &~ 0
\\
~0 &~ 0 &~m_0
\end{array}\hs{-0.15cm}\right)+ }
\end{array}  
\hs{-0.5cm}
\begin{array}{cc}
& {\begin{array}{ccc}
\hs{-2cm}~ &~  &~
\end{array}}\\ \vspace{2mm}
\begin{array}{c}
  \\ \\ 

\end{array} &{\hs{-0.4cm}\left(\begin{array}{ccc}
1-\ga &~~\fr{2\tl{\al }-\de |t|}{\sq{2+|t|^2}}&~~
\fr{\sq{2}(\tl{\al }|t|+\de )}{\sq{2+|t|^2}}
\\
\fr{2\tl{\al }-\de |t|}{\sq{2+|t|^2}} &~~
\fr{2\tl{\bt }+|t|^2(1+\ga )}{2+|t|^2}&~~
\fr{\sq{2}|t|}{2+|t|^2}(\tl{\bt 
}\hs{-0.1cm}-\hs{-0.1cm}1\hs{-0.1cm}-\hs{-0.1cm}\ga )
\\
\fr{\sq{2}(\tl{\al }|t|+\de )}{\sq{2+|t|^2}}&~~
\fr{\sq{2}|t|}{2+|t|^2}(\tl{\bt 
}\hs{-0.1cm}-\hs{-0.1cm}1\hs{-0.1cm}-\hs{-0.1cm}\ga )&~~
\fr{\tl{\bt }|t|^2+2(1+\ga )}{2+|t|^2}
\end{array}\hs{-0.2cm}\right)m_0\ep  ~.
}
\end{array}
\label{totenu1}
\end{equation}     
{}From (\ref{totenu1}) one can see that the additional rotations that
are needed to diagonalize the neutrino mass matrix can yield a small
$\te_{13}$ ($\tan \te_{13}\sim \ep $).  
Also, the correction to the 2-3 rotation is suppressed by $\ep $.
However, the 1-2
rotation can be naturally large. The matrix $m_{\nu }'$ is diagonalized by 
\beq 
(U_{\nu }^{(1)})^Tm_{\nu }'U_{\nu }^{(1)}= 
{\rm Diag } (m_1 ,~m_2 ,~ m_0)
\la{transnu1}
\eeq
with 
\begin{equation}
\begin{array}{ccc}
 & {\begin{array}{ccc}
 & & 
\end{array}}\\ \vspace{2mm}
U_{\nu }^{(1)}\simeq 
\begin{array}{c}
\end{array}\!\!\!\!\!\! &{\left(\begin{array}{ccc}
\hs{-0.1cm}c_{\te }~&s_{\te } ~ &|\ka |\ep
\\
\hs{-0.2cm}-s_{\te }e^{{\rm i}\phi }~ &c_{\te }e^{{\rm i}\phi }~& 
|\ka'| \ep e^{{\rm i}(\phi_{\ka }-\phi_{\ka' }) }
\\
\hs{-0.1cm}-(\ka c_{\te }-\ka's_{\te }e^{{\rm i}\phi })\ep ~&
-(\ka s_{\te }+\ka'c_{\te }e^{{\rm i}\phi })\ep ~&e^{{\rm i}\phi_{\ka' }}
\end{array} \right)\! }~,
\end{array}  
\label{Unu1}
\end{equation} 
where
$c_{\te }\equiv \cos \te $, $s_{\te }\equiv \sin \te $ and
\beq 
\tan 2\te =-\fr{2|b|}{|a-ce^{{\rm i}2\phi }|}~,~~~
\ka =\fr{\sq{2}(\tl{\al }|t|+\de )}{\sq{2+|t|^2}}~,~~~
\ka' =\fr{\sq{2}|t|}{2+|t|^2}(\tl{\bt }-1-\ga )~,
\la{angles}
\eeq
with
\beq
a=1-\ga ~,~~~
b=\fr{2\tl{\al }-\de |t|}{\sq{2+|t|^2}} ~,~~~
c=\fr{2\tl{\bt }+|t|^2(1+\ga )}{2+|t|^2} ~,~~~
\phi_{\ka , \ka' }={\rm Arg}(\ka , \ka' )~.
\la{abc}
\eeq
The phase $\phi $ is determined by
\beq
\sin (\phi -\phi_b+\phi_c)=
\left | \fr{a}{c}\right | \sin (\phi +\phi_b+\phi_a)~,~~~~
{\rm where}~~~~\phi_{a, b, c}={\rm Arg}(a, b, c)~. 
\la{phi}
\eeq
The masses of the three neutrino eigenstates are\footnote{In general,
$m_{1,2,3}$ are complex and an additional diagonal phase matrix is needed
to make them real. However, this phase matrix does not play a role in 
neutrino oscillations and can be ignored.} 
$$ 
m_1\simeq \lq ac_{\te
}^2-2bs_{\te
}c_{\te }e^{{\rm i}\phi } +cs_{\te }^2e^{{\rm i}2\phi }\rq m_0\ep ~,
$$
$$
m_2\simeq \lq as_{\te }^2+2bs_{\te }c_{\te }e^{{\rm i}\phi } 
+cc_{\te }^2e^{{\rm i}2\phi }\rq m_0\ep ~,
$$
\beq
m_3\simeq m_0~.
\la{numasses}
\eeq
{}From (\ref{numasses}),
$$
\De m_{\rm atm}^2=|m_3|^2-|m_2|^2\simeq |m_0|^2~,
$$
\beq 
\De m_{\rm sol}^2=|m_2|^2-|m_1|^2\sim |m_0|^2\ep^2~.
\la{massq}
\eeq
Therefore, the estimated value of $\ep $ is
\beq
\ep \sim \sq{\fr{\De m_{\rm sol}^2}{\De m_{\rm atm}^2}}=0.13-0.23~.
\la{epsval}
\eeq

Now let us discuss the lepton mixing angles.
Upon its full diagonalization the neutrino mass matrix was transformed as
\beq
(m_{\nu })^{diag}=U_{\nu }^Tm_{\nu }U_{\nu }~,
\la{diagmnu}
\eeq
where
\beq
U_{\nu }=U_{\nu }^{(0)}U_{\nu }^{(1)}~.
\label{Unu}
\end{equation} 
The lepton mixing matrix is given by
\beq
(V^l)_{\al i}=(U_e^{\dagger }U_{\nu })_{\al i}~,
\la{Vl}
\eeq
where $\al $ refers to the flavor index ($\al =e, \mu , \tau $) and 
$i=1, 2, 3$ denotes the light neutrino mass eigenstates.
One finds
\begin{equation}
\begin{array}{ccc}
 & {\begin{array}{ccc}
 & & 
\end{array}}\\ \vspace{2mm}
V^l_{\al i}\simeq 
\begin{array}{c}
\end{array}\!\!\!\!\!\! &{\left(\begin{array}{ccc}
c_{\te }~,&s_{\te }~,&|\ka |\ep 
\\
-\fr{2e^{{\rm i}\om }+|t\rho |}{\sq{(2+|t|^2)(2+|\rho |^2)}}s_{\te }
e^{{\rm i}\phi }~, &
\fr{2e^{{\rm i}\om }+|t\rho |}{\sq{(2+|t|^2)(2+|\rho |^2)}}c_{\te }
e^{{\rm i}\phi }~,&
\fr{\sq{2}(|t|e^{{\rm i}\om }-|\rho |)}{\sq{(2+|t|^2)(2+|\rho |^2)}}
e^{{\rm i}\phi_{\ka } }
\\
\fr{\sq{2}(|t|-|\rho |e^{{\rm i}\om })}
{\sq{(2+|t|^2)(2+|\rho |^2)}}s_{\te }e^{{\rm i}\phi }~,&
-\fr{\sq{2}(|t|-|\rho |e^{{\rm i}\om })}
{\sq{(2+|t|^2)(2+|\rho |^2)}}c_{\te }e^{{\rm i}\phi }~,
 &\fr{2+|t\rho |e^{{\rm i}\om } }{\sq{(2+|t|^2)(2+|\rho |^2)}}
e^{{\rm i}\phi_{\ka } } 
\end{array} \right)_{\al i}\! }~,
\end{array}  
\label{Vlexp}
\end{equation} 
where $\om ={\rm Arg }(\rho )-{\rm Arg }(t)$.
{}From (\ref{Vlexp}), we find 
\beq
\tan \te_{\mu \tau }\simeq \sq{2}
\left |\fr{t-\rho }
{2+t\rho^* }\right |~,~~~~
\tan \te_{e \mu ,\tau }\simeq \tan \te ~.
\la{atmsolpar}
\eeq
According to the democratic approach, we have $t, \rho , \tan \te \sim 1$,
so that
\beq
\sin^22\te_{\mu \tau}\sim 1~,~~~~~
\sin^22\te_{e \mu ,\tau}\sim 1~.
\la{bilarge}
\eeq
Thus, bilarge neutrino mixing is realized.
{}For the remaining angle $\te_{13}$, 
from (\ref{Vlexp}) taking (\ref{epsval}) into account,
\beq
\te_{13}\equiv V^l_{e3}\simeq |\ka |\ep \sim 
|\ka |\sq{\fr{\De m_{\rm sol}^2}{\De m_{\rm atm}^2}}~.
\la{Vle3}
\eeq
Thus,  $\te_{13}$ is naturally suppressed ($\sim \ep $), but seems within
reach of the next round experiments.

\section{Conclusions}

By exploiting a democratic approach supplemented by suitable symmetries, we
have attempted to obtain an understanding of the charged lepton mass
hierarchies, bilarge neutrino mixing as well as a suppressed mixing angle
$\te_{13}$.
Since we studied this  in the MSSM framework 
augmented with singlet states, the lepton sector does not have any impact
on quark masses and their mixings. Because of this, the quark sector can
also nicely blend with neutrino democracy. However, this will change if
some GUT scenario such as $SU(5)$ and $SO(10)$ is considered, 
where the quark and lepton mass
matrices can be related to each other. For realistic
pattern of fermion masses and mixings some extensions will become
necessary \cite{ourdemo1}, \cite{ourdemo2}.

\vs{0.5cm}

\hs{-0.7cm}{\bf Acknowledgments}

\vs{0.2cm} 
\hs{-0.7cm}We would like to acknowledge the hospitality of CERN theory
division, where part of this work was done.
Q.S. acknowledges the hospiality of the Institute of Theoretische
Physik (Heidelberg), especially Michael Schmidt and Christof Wetterich,
as well as the Alexander von Humboldt Stiftung.
This work is supported by NATO Grant PST.CLG.977666 and
by DOE under contract DE-FG02-91ER40626.


\bibliographystyle{unsrt}

\end{document}